\documentclass{article}
\usepackage{dcase2020,amsmath,graphicx,url,times,booktabs, tabularx}
\usepackage{tablefootnote}
\usepackage{comment}
\usepackage{mathtools}
\usepackage{amssymb}
\usepackage{csquotes}
\usepackage{multirow}
\usepackage{hyperref}
\usepackage{comment}
\usepackage{float}

\usepackage{enumitem}
 \usepackage[table,xcdraw]{xcolor}

\hypersetup{
    colorlinks=true,
    linkcolor=blue,
    filecolor=magenta,      
    urlcolor=cyan,
}

\title{Low-complexity CNNs  for Acoustic scene classification}

\name{Arshdeep Singh, Mark D. Plumbley}
\address{Centre for Vision, Speech and Signal Processing (CVSSP) \\ University of Surrey, UK\\
Email: $\{arshdeep.singh,m.plumbley\}$@surrey.ac.uk}

\begin{document}

\ninept
\maketitle

\begin{sloppy}
\begin{abstract}

This paper presents a low-complexity framework for acoustic scene classification (ASC). Most of the frameworks designed for ASC use convolutional neural networks (CNNs) due to their learning ability and improved performance compared to hand-engineered features. However, CNNs are resource hungry due to their large size and high computational complexity. Therefore, CNNs are difficult to deploy on resource constrained devices. This paper addresses the problem of reducing the computational complexity and memory requirement in CNNs. We propose a low-complexity  CNN architecture, and apply pruning and quantization to further reduce the parameters and memory. We then propose an ensemble framework that combines various low-complexity CNNs to improve the overall performance. An experimental evaluation of the proposed framework is performed on the publicly available DCASE 2022 Task 1 that focuses on ASC. The proposed ensemble framework  has approximately 60K parameters, requires 19M multiply-accumulate operations and  improves the performance by   approximately 2-4 percentage points compared to the  DCASE 2022 Task 1 baseline network.


\end{abstract}


\section{Introduction}
\label{sec:intro}

Convolutional neural networks (CNNs) have shown  state-of-the-art performance  in comparison to traditional hand-crafted methods in various domains \cite{gu2018recent}.  However, CNNs are resource hungry due to their large size and  computational complexity \cite{simonyan2014very, krizhevsky2012imagenet}, and hence it is difficult to deploy CNNs on resource constrained devices. For example, Cortex-M4 devices (STM32L496@80MHz or Arduino Nano 33@64MHz) have a  maximum allowed limit of 128K parameters and 30M multiply-accumulate operations (MACs) per second during inference. Thus, the issue of reducing the size  and the computational cost of CNNs has drawn a significant amount of attention in the detection and classification of acoustic scenes and events (DCASE) research community.


In the literature, some  CNN parameters such as filters or weights may be redundant, and contribute to extra memory and computational complexity only \cite{wen2020exploiting,singh2020svd}.
For example, Li et al. \cite{li2016pruning} found  that 64\% of the parameters  which contribute  approximately 34\% of computation time are redundant. Removing the redundant parameters from CNNs gives similar performance with an advantage of reduced memory and less computational cost.

The majority of the methods applied to eliminate redundant parameters are on filter pruning \cite{liu2017learning, he2017channel,he2019filter,lin2020hrank}, where a redundant filter from the network is being eliminated. Filter pruning methods have been widely employed in the computer vision. However, only a few works \cite{singh2020svd,kim2021qti} have applied filter pruning methods in the audio domain, and the issue of designing CNNs for resource constrained devices with constraints  on both memory and MACs have not yet been fully explored.   Recently, the DCASE community has considered on designing low-complexity frameworks for ASC with constraints on both memory and MACs. The  DCASE challenge 2022 Task 1 \cite{martin2022low} provides a baseline CNN which has 46512 parameters and 29.24M MACs with accuracy approximately 43\% and 1.575 log-loss for audio scene classification (ASC) .

This paper aims to design \enquote{low-complexity CNNs} which have a maximum number of parameters less than 128K and a maximum number of MACs per seconds less than 30M and performance better than that of the DCASE 2022 Task 1 baseline network for ASC.


The major contributions of the paper is summarized below,
\begin{enumerate}[label=(\alph*)]
    \item We design a \enquote{low-complexity} CNN that has a fewer parameters and MACs with  better performance  than that of the DCASE 2022 Task 1 baseline CNN. 
    \item A filter pruning method is applied to compress the \enquote{low-complexity CNN} of (a) further. Subsequently, we quantize  each parameter from float32 to INT8 data type, reducing networks memory by four times.
    \item An ensemble approach is proposed which combines predictions obtained from several (b) low-complexity CNNs.
    \item Experimental evaluation is undertaken to compare  performance of classical methods such as Gaussian Mixture Model (GMMs) and random forest (RF) classifiers, learning based methods such as dictionary learning and a pre-trained high complexity CNN, VGGish with the proposed framework.
\end{enumerate}



The rest of the paper is organized as follows. 
In Section \ref{sec: dataset used}, a brief overview of dataset used and  features used for experimentation is described. In Section  \ref{sec: low-complexity CNNs}, a procedure to obtain low-complexity CNN and an ensemble framework  is described. Section \ref{sec: perfromance}  presents  experimental analysis. Finally, discussion and conclusion is presented in Section \ref{sec: DISCUS} and \ref{sec: conclusion} respectively.

\section{Experimental dataset and feature extraction}
\label{sec: dataset used}
DCASE 2022 Task 1 uses the TAU Urban Acoustic Scenes 2022 Mobile, development and evaluation datasets \cite{Heittola2020}. The dataset contains recordings from 12 European cities in 10 different acoustic scenes using 4 different recording devices. Each audio recording has 1 second length. The development dataset is divided in training and validation sets. The training dataset consists of  139620 audio examples and the validation dataset consists of 29680 audio examples. The evaluation dataset consists of only audio examples without any groundtruth available publicly, and the evaluation is performed by the DCASE challenge community.

\noindent \textbf{Feature extraction:} For time-frequency representations, log-mel band energies of size (40 $\times$ 51) corresponding to an audio signal of 1 second length are extracted. A Hamming asymmetric window of length  40ms, and a hop length of 20ms is used to extract magnitude spectrogram. Next, log-mel spectrogram is computed using 40 mel bands.

\section{Obtaining low-complexity CNNs}
\label{sec: low-complexity CNNs}
\noindent \textbf{Low-complexity optimal CNN architecture:} We design a simple low-complexity CNN architecture which  consists of three convolutional layers (C1, C2 \& C3), two pooling layers (P1, P2), a dense layer (D) and a classification layer.  We perform an empirical analysis to select an appropriate size of CNN filters and activation functions across the different layers. The filter size for each convolutional layer is chosen from a set, $\{1 \times 1,  3 \times 3, 5 \times 5, 7 \times 7\}$. We choose hyperbolic tangent (tanh) or rectified linear unit (ReLU) activation function for different layers. 
The low-complexity  CNN is trained using the training dataset with a batch size of 64 with an Adam optimizer for 1000 epochs. A categorical cross-entropy loss function is used during the training process. We apply an early stopping criterion to yield the best network that gives the minimum log-loss for the validation dataset.

The optimal architecture obtained after performing empirical analysis is given in Table \ref{tab: low-complxity architecutre}. The details of the experiments are given in Section \ref{sec: perfromance}. The proposed architecture requires approximately 5M MACs to produce an output corresponding to an input of size  (40 x 51), and has 14886 parameters. The performance of the trained network is measured in terms of accuracy and log-loss, averaged over 5 different iterations.

\noindent \textbf{Reducing redundancy in the  low-complexity optimal CNN via filter pruning:} To eliminate redundant filters from the low-complexity optimal architecture as given in Table \ref{tab: low-complxity architecutre}, we apply a filter pruning strategy. For each convolutional layer, we identify filter pairs which are similar. Our hypothesis is that similar filters produce similar output or feature maps and hence, contribute to redundancy only. Therefore, one of the similar filters can be eliminated. The similarity between the filters is measured using a cosine distance. We identify the closest filter pairs for each layer separately. A filter from each pair is deemed redundant and eliminated from the network. More information about the similarity based  filter pruning method can be found at  \cite{singh2022passive}.  

The number of redundant filters obtained after performing  similarity-based filter pruning for C1 layer is 4 out of 16, C2 layer is 4 out of 16 and C3 layer is 10 out of 32. We obtain 6 different pruned networks that are obtained after pruning C1 layer only, C2 layer only, C3 layer only, C1 and C2 layers, C2 and C3 layers, C1 and C2 and C3 layers. The number of MACs and the number of parameters for each pruned network are given in Table \ref{tab:  various CNNs DCASE2022}.

To regain the loss in performance due to pruning, the pruned networks are fine-tuned with a similar process to that used for the training of the unpruned low-complexity optimal network.

\noindent \textbf{Reducing memory requirement via quantization:} To reduce size of the network, we  perform quantization on parameters of each pruned network using  Tensorflow-Lite (TFLite). TFlite \cite{david2021tensorflow} is  an open-source framework to quantizes a pre-trained full-precision network for embedded devices. 
We quantize the network parameters from 32-bit floating point  to 8-bit integers. This leads to reduce the network size 4x.


\noindent \textbf{An ensemble framework:} Next,  the predictions obtained from various pruned networks are aggregated together in an ensemble framework. The total number of parameters in the ensemble framework that aggregates predictions from all 6 pruned networks are 70.97K and the total number of MACs are 23.84M.

The codes and the pruned networks can be found online \footnote{\href{https://github.com/Arshdeep-Singh-Boparai/DCASE2022_Challenge.git}{Link: Pruned Quantized models, confusion matrices, evaluation scripts.}}\footnote{\href{https://github.com/AlbertoAncilotto/NeSsi}{Link: Model size and complexity calculation.}}.

\begin{table}[t]
\caption{Low-complexity optimal CNN architecture. Here, tanh is a hyperbolic tangent activation function and ReLU is a rectified linear unit activation function.}
\vspace{0.15cm}
\Large
\hspace{0.32cm}
\label{tab: low-complxity architecutre}
\resizebox{0.4\textwidth}{!}{
\begin{tabular}{ccc}
\toprule
Layer name & Description                       \\		\midrule
{C1} & {\begin{tabular}[c]{@{}l@{}}Convolution 16@(3 $\times$ 3) +  Batch Normalization + tanh\end{tabular}}                                         \\ \midrule
{C2} & {\begin{tabular}[c]{@{}l@{}}Convolution 16@(3 $\times$ 3) + Batch Normalization + ReLU\end{tabular}}                                       \\ \midrule
{P1} & {Average Pooling (5 $\times$ 5)}                                   \\\midrule
{C3} & {\begin{tabular}[c]{@{}l@{}}Convolution 32@(3 $\times$ 3) + Batch Normalization + tanh\end{tabular}}                                     \\\midrule
{P2} & {Average Pooling  (4 $\times$ 10)}   \\ \midrule
{D} &{\hspace{-1.5cm}\begin{tabular}[c]{@{}l@{}}Dense (100) + tanh \end{tabular}}                                         \\ 		\midrule
{Classification} & {\hspace{-1.5cm} softmax (10)}                                \\   
\bottomrule
\end{tabular}}
\end{table}

\begin{table*}[h]
	\caption{Various low-complexity CNNs obtained after pruning and applying quantization (INT8).}
	\centering
	\label{tab:  various CNNs DCASE2022}
	\vspace{0.05cm}
	\Large
	\resizebox{0.85\textwidth}{!}{
	\begin{tabular}{ccccccc} \toprule
		{Sr No.}& {Network Name} & {Pruned layer}  & {Architecture (C1-C2-C3-Dense)} & {Parameters} & {Size (KB)} & {MACs (millions)}  \\
		\midrule
		{1} &{Unpruned optimal low-complexity} & {NA} &{16-16-32-100}& {14886} & { 18.59 } & {5.41}  \\
		\midrule
		{2}& {Pruned\_C1} &{C1} & {12-16-32-100} &{14254} & { 17.86 } & {4.16}  \\
		\midrule
		{3}&{Pruned\_C2} & {C2} & {16-12-32-100} & {13138} & { 16.85} & {4.13} \\
		\midrule
		{4} & {Pruned\_C3} &{C3} & {16-16-22-100} & {11396} & { 15.11 } & {5.29} \\
        \midrule
		{5} & {Pruned\_C12} &{C1 + C2} & {12-12-32-100} & {12650} & { 16.26 } & {3.18} \\
		\midrule
		{6}&{Pruned\_C23} &{C2 + C3} & {16-12-22-100} &{10008} &  { 13.73} & {4.04}  \\
		\midrule
		{7}& {Pruned\_C123} &{C1 + C2 + C3} & {12-12-22-100} & {9520} & { 13.14 } & {3.08} \\
	    \bottomrule
	\end{tabular}}
\end{table*}

\section{Performance Analysis}

\label{sec: perfromance}

The accuracy and the log-loss obtained at various filter size and different activation functions in the low-complexity CNN is shown in Figure \ref{fig: filter_selection}. The low-complexity CNN with (3 x 3) filter size  outperforms networks with other filter size.

Using  \enquote{ReLU} activation function across all layers results in smaller log-loss in comparison to that of the  the \enquote{tanh} activation function. On the other hand, combining \enquote{tanh} and \enquote{ReLU} activation functions across different layers improves the log-loss further as shown in Figure \ref{fig: activation_comparison}. We find that the  \enquote{tanh} activation function in the first layer and \enquote{ReLU} for other layers improves the performance over that of using \enquote{ReLU} for all layers. Finally, we choose the  low-complexity optimal CNN architecture which performs better than that of architectures with different combinations of activation function. The optimal low-complexity CNN has \enquote{tanh} activation function in all layers except C2 and the filter size is (3 x 3) in each convolutional layers.

\begin{figure}[h]
    \centering
    \includegraphics[scale=0.32]{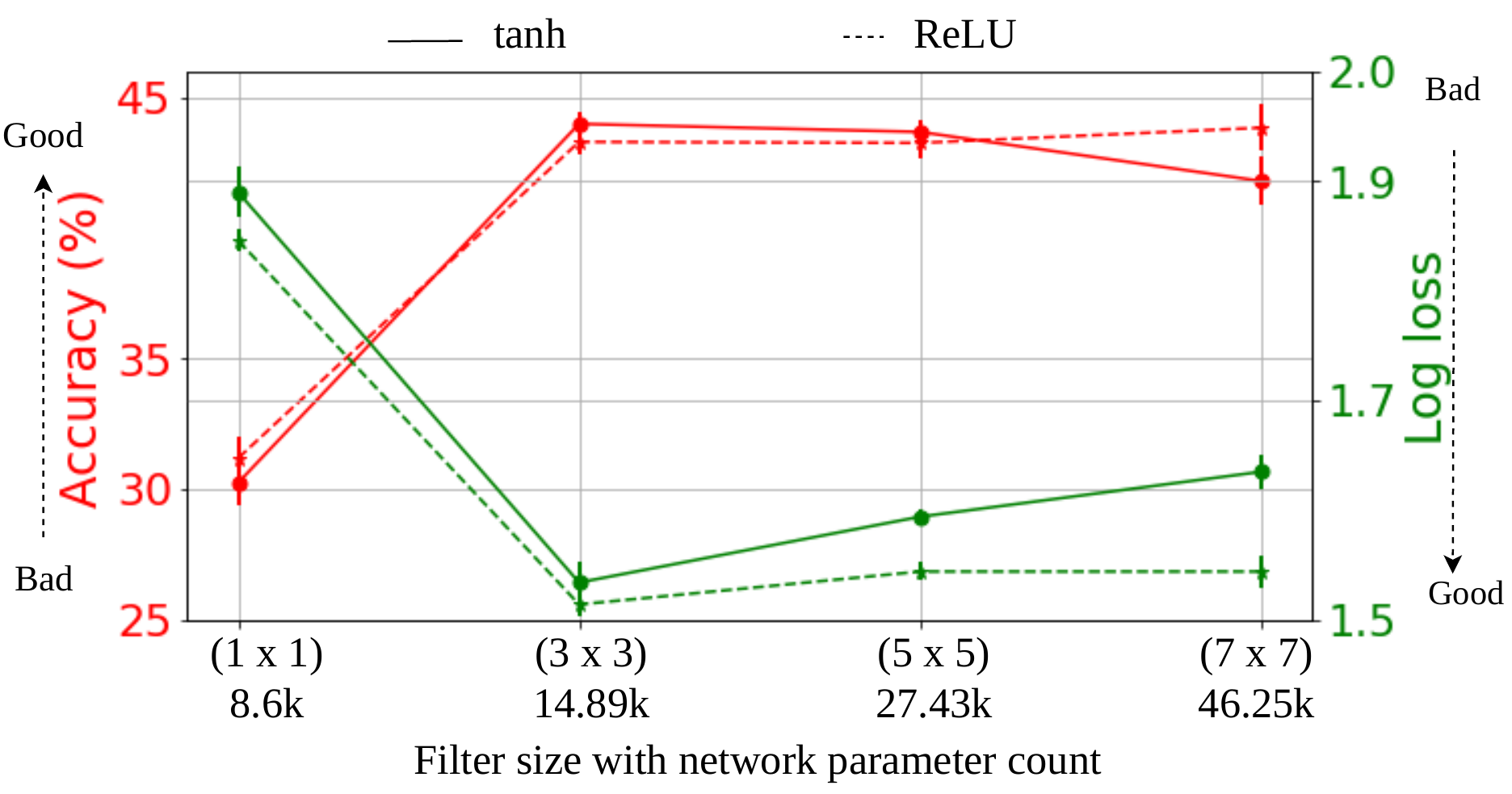}
    \vspace{-0.5cm}
    \caption{Accuracy and log loss obtained for  DCASE 2022 Task 1 development validation dataset using the low-complexity CNNs at different filter size and activation function.}
    \label{fig: filter_selection}
\end{figure}

\begin{figure}[h]
    \centering
    \includegraphics[scale=0.45]{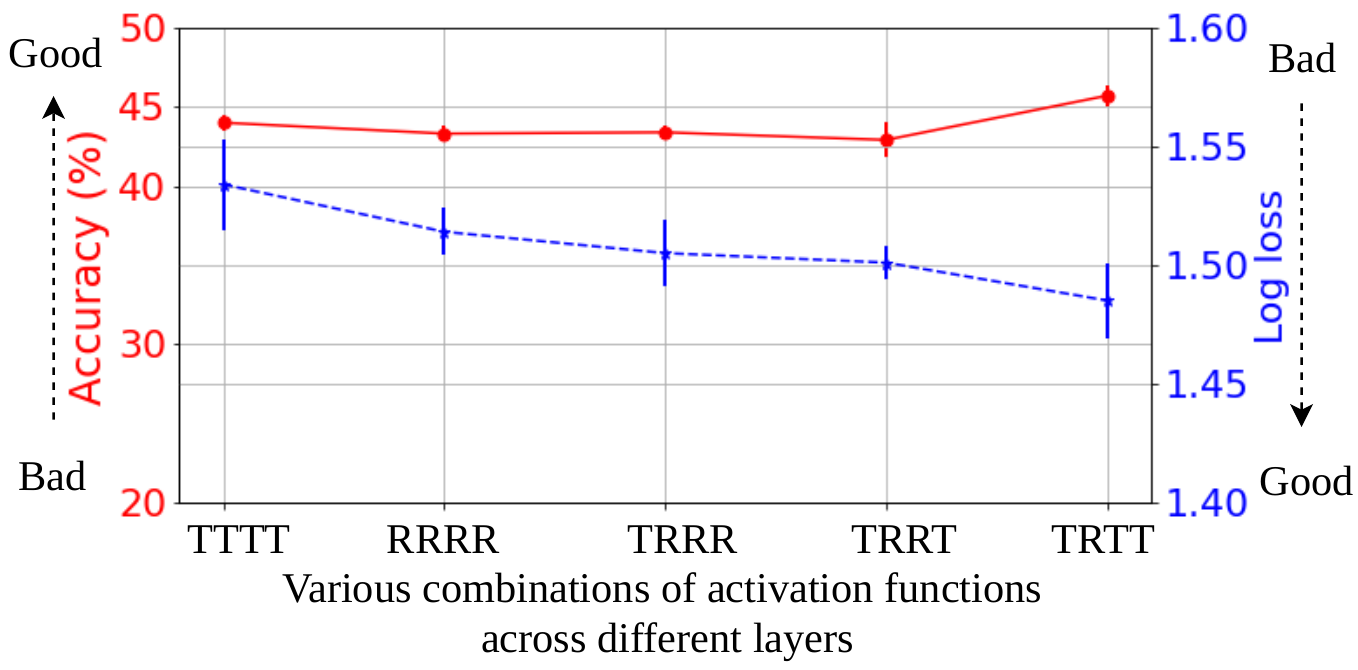}
    \vspace{-0.5cm}
    \caption{Accuracy and log loss obtained  for DCASE 2022 Task 1 development validation dataset using the low-complexity CNN with (3 x 3) filter size when different combinations of activation functions are applied across \enquote{C1}, \enquote{C2}, \enquote{C3} and \enquote{D} layers. Here, \enquote{T} represents tanh activation function and \enquote{R} represents ReLU activation function.}
    \label{fig: activation_comparison}
\end{figure}

Next, we analyse the performance obtained using the  low-complexity optimal CNN without performing any pruning and after pruning it. Figure \ref{fig: performance of the pruned network} shows the accuracy and the log-loss obtained for the unpruned  low-complexity optimal network and its various pruned networks. The unpruned low-complexity optimal network gives 1.475 log-loss and 45.92\% accuracy.  Eliminating filters from the unpruned  network 
results in a significant reduction in performance, but this is almost entirely restored after fine-tuning.

The performance obtained after aggregating predictions from various pruned networks is given in Table \ref{tab:  ensemble DCASE2022}. The ensemble framework improves the performance in comparison to that of individual pruned networks.

\noindent \textbf{Performance comparisons:}  
The proposed ensemble framework improves performance as compared to the DCASE 2022 Task 1 baseline network for development and evaluation datasets as given in Table  \ref{tab: performance comaprison_dcase}. We also compare the performance with the following methods,

\begin{itemize}

    \item \textbf{GMM:} We train Gaussian Mixture Models (GMMs) for each scene class separately, and perform classification using maximum likelihood estimates. The log-mel spectrogram of size (40 x 51) is averaged along temporal dimension to yield (40 x 1) vector, which is given as an input to train the GMMs. The number of mixtures (n) per class are chosen from \{5,10,15,20,30,50\}. 
    
    \item \textbf{RF:} A random forest (RF) classifier is trained using the log-mel spectrogram averaged along the temporal dimension. The number of estimators are set to 100 and the depth (d) is chosen from \{5,20,40,50,100,200\}.
    
    \item \textbf{Dictionary learning}:  A dictionary learning framework with structured incoherence and shared features (DLSI) \cite{ramirez2010classification} is trained. DLSI framework learns the dictionary for each class by minimizing the reconstruction error and  reduces the redundant dictionary atoms in the learning process itself.  The number of dictionary atoms per class (k) are chosen from \{5,10,15,20,40\}.
    
    \item \textbf{MLP}: A multi-layer perceptron (MLP) network with 2 dense layers each having 100 and 50 units is trained on averaged log-mel spectrogram along the temporal dimension.
    
    \item \textbf{Pre-trained CNN}: 
    We use pre-trained convolutional layers of VGGish \cite{cramer2019look,VGGish} followed by a dense and a classification layer to yield an end-to-end VGGish network. The input to the end-to-end VGGish is log-mel spectrogram of size (40 x 51). We train the end-to-end VGGish for 1000 epochs with similar training settings as used to train the low-complexity CNNs.
    
\end{itemize}

The proposed ensemble framework  outperforms the other methods as shown in  Figure \ref{fig: comparison other methods}. It is interesting to note that the proposed low-complexity optimal CNN  performs better than that of the large-scale pre-trained CNN which has 4.5M parameters and has 1077M MACs.





\begin{figure}[h]
    \centering
    \includegraphics[scale=0.34]{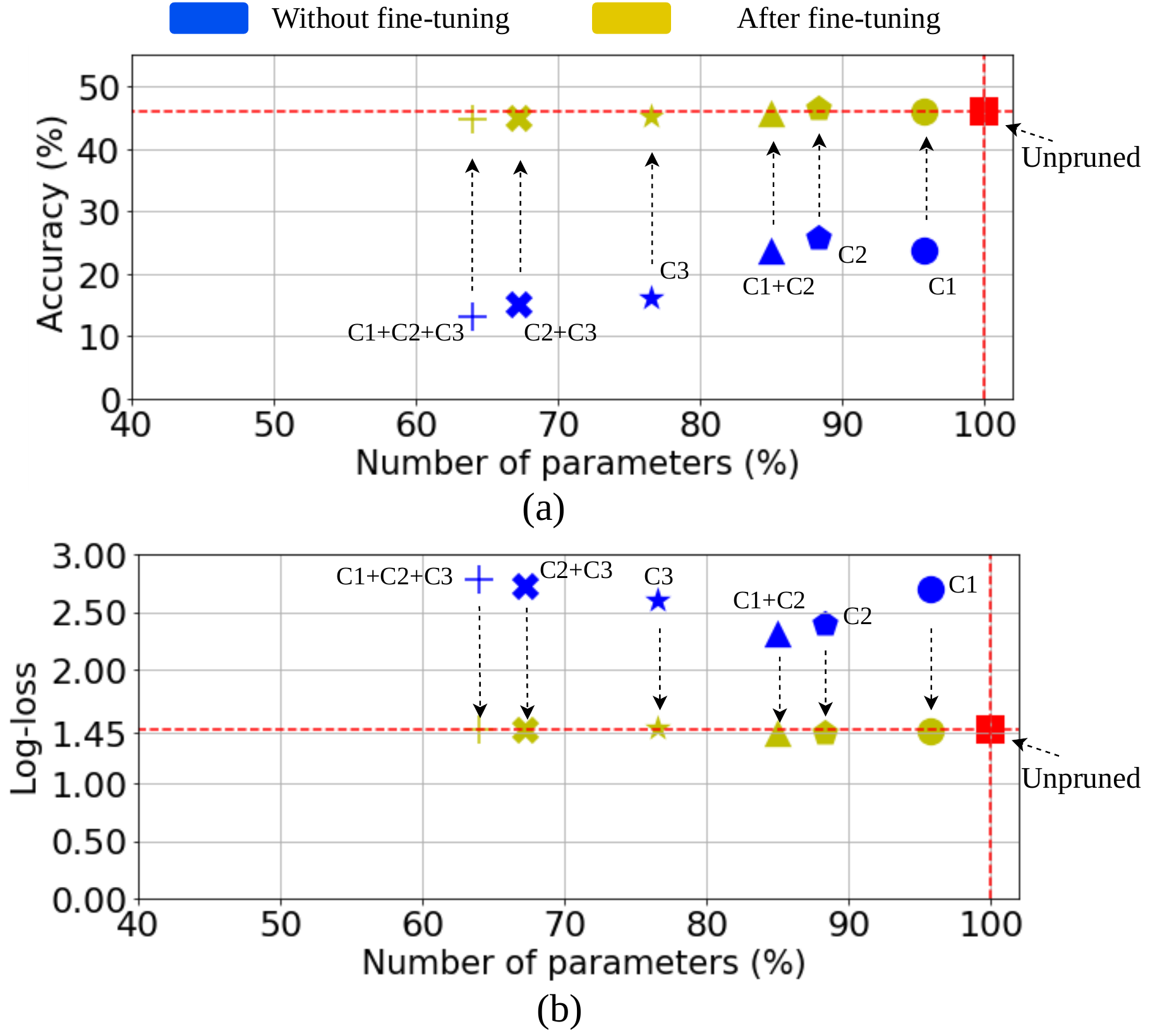}
    \vspace{-0.5cm}
    \caption{(a) Accuracy and (b) log-loss obtained after pruning different intermediate layers (C1, C2, C3, C1+C2, C2+C3, C1+C2+C3) in the unpruned low-complexity optimal CNN for  DCASE 2022 Task 1 development validation dataset. The accuracy and the log-loss is obtained with and without performing the fine-tuning of the pruned network.}
    \label{fig: performance of the pruned network}
\end{figure}

\begin{figure}[h]
    \centering
    \includegraphics[scale=0.43]{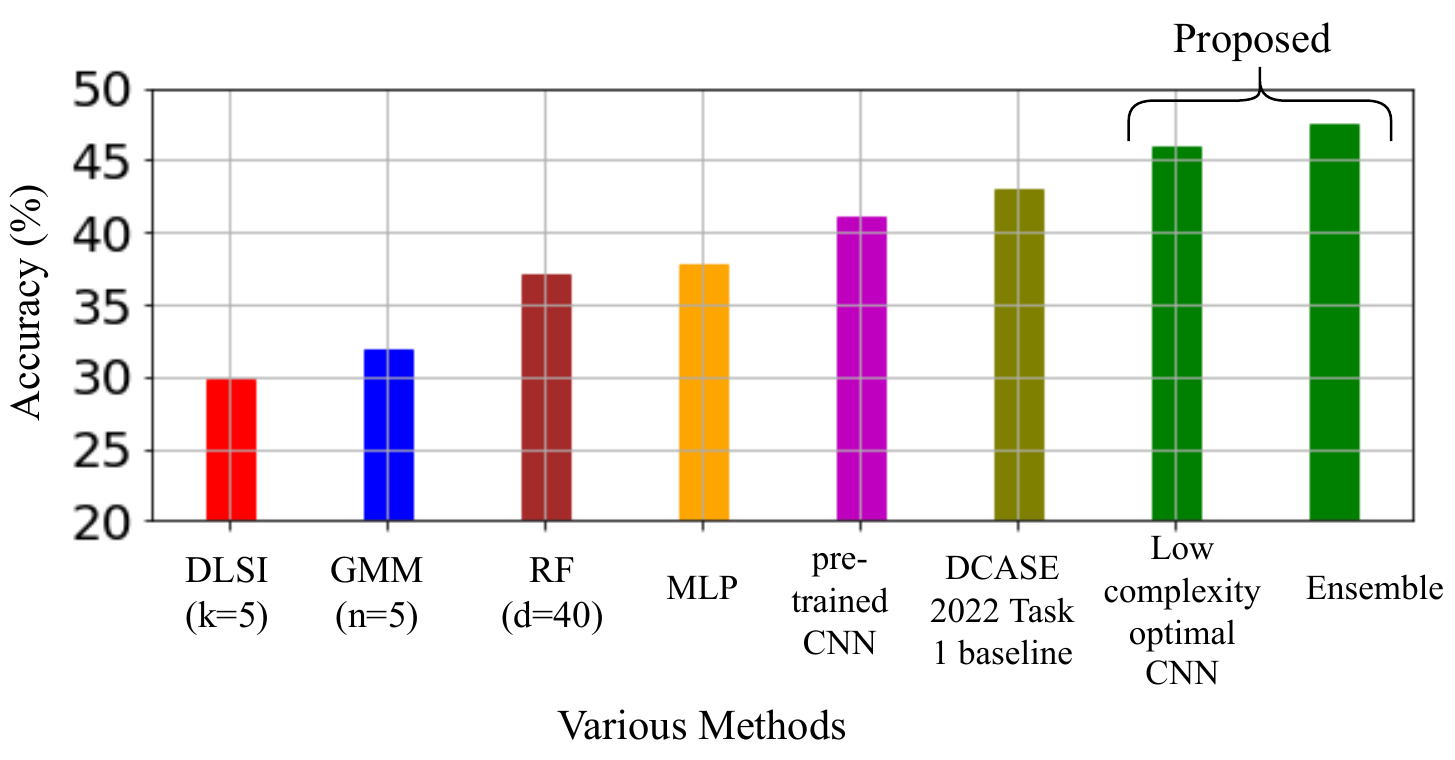}
    \vspace{-0.5cm}
    \caption{Accuracy comparison of various methods for DCASE 2022 Task 1 development validation set. Here, we show the best accuracy obtained from the DLSI, GMM and RF framework.}
    \label{fig: comparison other methods}
\end{figure}

\begin{figure}[t]
    \includegraphics[scale=0.35]{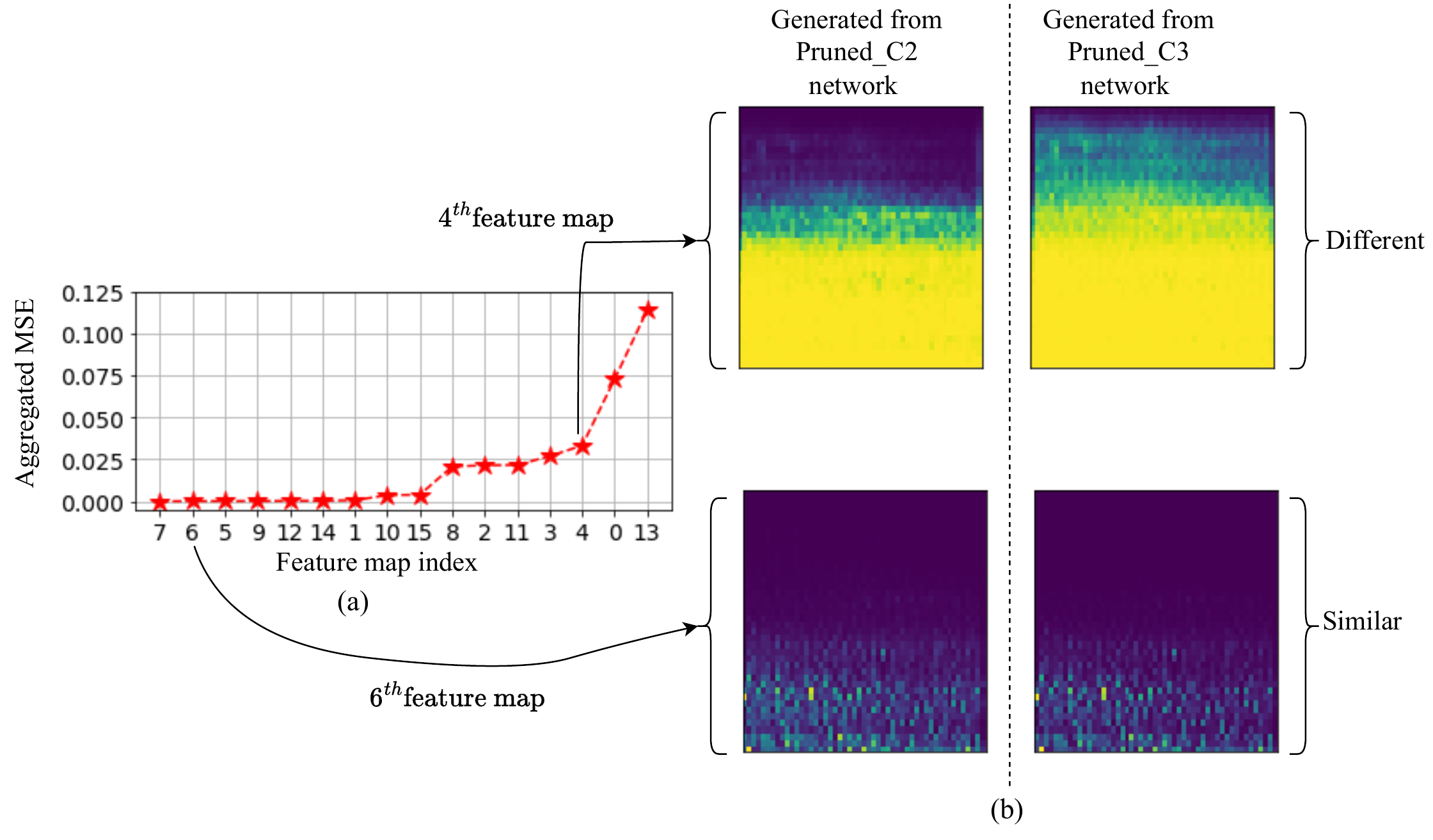}
    \vspace{-0.5cm}
    \caption{(a) Aggregated MSE computed between the feature maps generated using Pruned\_C2 and Pruned\_C3 network from the C1 layer. (b) shows  the  different and  the similar feature maps corresponding to $4^{\text{th}}$ and $6^{\text{th}}$ feature map index.}
    \label{fig: ensemble redundancy}
\end{figure}

\begin{table*}[t]
	\centering
	\caption{Performance obtained for  DCASE 2022 Task 1 development validation dataset using the ensemble framework that combines various low-complexity optimal CNNs obtained after pruning and quantization.}
	\Large
	\label{tab:  ensemble DCASE2022}
	\vspace{0.1cm}
	\resizebox{0.9\textwidth}{!}{
	\begin{tabular}{cccccccc} \toprule
		{Sr No.}& {Ensemble framework} & {Number of parameters}  & {MACs (millions)} & {Size (KB)} & {Accuracy (\%)} & {Log-loss}  \\
		\midrule
		{1}& {All pruned networks except Pruned\_C1} &{56712} & {19.72} & {75.09 } & {47.14} & {1.394}  \\
		\midrule
		{2}&{All pruned networks except Pruned\_C2} & {57828} & {19.75} & {76.10} & {47.10} & {1.396} \\
		\midrule
		{3} & {All pruned networks except Pruned\_C3} &{59570} & {18.60} & {77.84} & {47.26} & {1.392} \\
        \midrule
		{4} & {All pruned networks except Pruned\_C12} &{58316} & {20.70} & {76.69} & {47.45} & {1.394} \\
		\midrule
		{5}&{All pruned networks except Pruned\_C23} &{60958} & {19.84} & {79.22} & {47.52} & {1.389}  \\
		\midrule
		{6} & {All pruned networks except Pruned\_C123} &{61446} & {20.80} & {79.81} & {47.35} & {1.392} \\
				\midrule
		{7} & {Ensemble on all pruned networks} & {70966} & {23.84} & {92.95} & {47.45} & {1.389} \\
       \bottomrule
	\end{tabular}}
\end{table*}


\begin{table}[h]
\caption{Performance comparison}
\label{tab: performance comaprison_dcase}
\resizebox{0.5\textwidth}{!}{%
\begin{tabular}{|c|cccc|}
\hline
\multirow{3}{*}{Framework} & \multicolumn{4}{c|}{Dataset}                                                                                       \\ \cline{2-5} 
                           & \multicolumn{2}{c|}{Development}                                   & \multicolumn{2}{c|}{Evaluation}               \\ \cline{2-5} 
                           & \multicolumn{1}{c|}{Accuracy (\%)} & \multicolumn{1}{c|}{log-loss} & \multicolumn{1}{c|}{Accuracy (\%)} & log-loss \\ \hline
\begin{tabular}[c]{@{}c@{}}DCASE baseline network \cite{martin2022low} \end{tabular} & \multicolumn{1}{c|}{42.9} & \multicolumn{1}{c|}{1.575} & \multicolumn{1}{c|}{44.2} & 1.532 \\ \hline
Proposed ensemble            & \multicolumn{1}{c|}{47.26}         & \multicolumn{1}{c|}{1.392}    & \multicolumn{1}{c|}{45.9}          & 1.492    \\ \hline
\end{tabular}%
}
\end{table}



\noindent \textbf{Similarity analysis among various pruned models:} We analyse similarity between the different pruned models as given in Table \ref{tab:  various CNNs DCASE2022}. For this, we use 10k audio examples from the validation dataset, and generated outputs from the various filters (feature maps). 
 The similarity is measured using the aggregated mean square error (MSE) across 10k examples, computed between the corresponding feature maps of two different pruned models.


We find that a few of the feature maps generated by the two different pruned models  are similar with MSE $\le 10^4$.  To show this, the aggregated MSE computed across corresponding feature maps generated by the   Pruned\_C2 and the Pruned\_C3 network is plotted in Figure \ref{fig: ensemble redundancy}(a). A  few of the feature maps have MSE close to zero, and hence these feature maps are  similar. We also show the different and the similar feature maps obtained from the Pruned\_C2 network and the Pruned\_C3 network corresponding to 4$^{\text{th}}$ and 6$^{\text{th}}$ index for a given input in Figure \ref{fig: ensemble redundancy}(b). This suggests that similar feature maps across different pruned networks are redundant, and are not required to compute again for the other network.  

We find that the Pruned\_C1 network shares 7 feature maps with the Pruned\_C12 network, and 3 feature maps with the Pruned\_C123 network across C1 layer. Similarly, the Pruned\_C2 network shares 7 feature maps with the Pruned\_C3, network and 6 feature maps with the Pruned\_C23. 

Ignoring similar feature maps across different pruned models, 
the MACs  are further reduced by 6M points, and the number of total parameters are reduced by approximately 3.4k in the ensemble framework that combines all pruned networks.

\section{Discussion}
\label{sec: DISCUS}
\noindent \textbf{Advantage of obtaining low-complexity CNNs via  Pruning:} The accuracy obtained using the Pruned\_C123 CNN  
having 9.5k parameters
is approximately 16 percentage point better than that of the low-complexity CNN with 8.6k parameters. 
This suggests that the smaller network obtained from the relatively large size network after pruning gives better performance as compared to the similar size smaller network.

We find that the ensemble on various CNNs improves the performance as compared to the individual CNN. However, the ensemble on the various CNNs consumes more resources. Therefore, pruning individual CNNs provide an advantage to use the ensemble framework efficiently. To improve the efficiency of the ensemble further, the shared feature maps (similarity) across various CNNs  can be ignored.

\section{Conclusion}
\label{sec: conclusion}
This paper focuses on designing a low-complexity system for acoustic scene classification. A filter pruning, quantization and ensemble procedure is applied to obtain compressed, accelerated, and low-size CNN. The proposed framework shows promising results in terms of reduction in  parameters and performance.  In future, our aim is to improve the performance of the low-complexity framework further.

\section{Acknowledgements}
This work was partly supported by Engineering and Physical Sciences Research Council (EPSRC) Grant EP/T019751/1 \enquote{AI for Sound (AI4S)}. For the purpose of open access, the authors have applied a Creative Commons Attribution (CC BY) licence to any Author Accepted Manuscript version arising.
Thanks to James A. King and Xubo Liu, PhD students at University of Surrey, UK for the discussions.


\newpage
\bibliographystyle{IEEEtran.bst}
\bibliography{refs.bib}

%
%
%
%
%
%
%
%
%

\end{sloppy}
\end{document}